\documentclass[useAMS,usenatbib]{mn2e}
\usepackage{amsmath,graphicx}
\usepackage[usenames]{color}
\usepackage{aas_macros}

\title[
Mode density and frequencies in HD~50844]
{Mode density and frequency extraction in $\delta$~Scuti star HD~50844}
\author[L. A. Balona]    	
{L. A. Balona
\\
South African Astronomical Observatory, P.O. Box 9, Observatory 7935, Cape Town, South
Africa}
\begin{document}

\date{Accepted .... Received ...}

\pagerange{\pageref{firstpage}--\pageref{lastpage}} \pubyear{2010}

\maketitle

\label{firstpage}

\begin{abstract}
We consider the high mode density reported in the $\delta$~Scuti star
HD~50844 observed by {\it CoRoT}.  Using simulations, we find that
extracting frequencies down to a given false alarm probability by means of
successive prewhitening leads to a gross over-estimate of the number of
frequencies in a star.  This is due to blending of the peaks in the
periodogram due to the finite duration of the time series.  Prewhitening is
equivalent to adding a frequency to the data which is carefully chosen to
interfere destructively with a given frequency in the data.  Since the 
frequency extracted from a blended peak is not quite correct, the
interference is not destructive with the result that many additional
fictitious frequencies are added to the data.  In data with very high
signal-to-noise, such as the {\it CoRoT} data, these spurious frequencies
are highly significant.  Continuous prewhitening thus causes a cascade of
spurious frequencies which leads to a much larger estimate of the mode
density than is actually the case.  The results reported for HD~50844
are consistent with this effect.  Direct comparison of the power in the raw
periodogram in this star with that in $\delta$~Scuti stars observed by {\it
Kepler} shows that HD~50844 has a typical mode density.
\end{abstract}


\section{Introduction}

The $\delta$~Scuti class of variables are dwarfs or giants with spectral types
between A2 and F5.  They lie on an extension of the Cepheid instability strip 
with periods in the range 0.02--0.3~d.  Most of the pulsational driving in 
these stars is by the $\kappa$~mechanism in the He{\sc II} partial  ionization 
zone.  A large number of  $\delta$~Sct stars have
been identified in photometric time series data from the {\it Kepler}
spacecraft.  A study of these stars by \citet{Balona2011c} has revealed that
even in the center of the instability strip, no more than half the stars
pulsate as $\delta$~Sct variables.  This is a surprising finding; the reason 
for damping of pulsations in constant stars in the instability strip is
presently not known.
 
Prior to the {\it MOST, CoRoT} and {\it Kepler} space missions, it was
thought that the frequency spectra in $\delta$~Sct stars may become very
dense as the detection threshold is lowered.  The reason is that modes of
high degree, $l$, which are not seen from the ground because of the low
amplitudes due to cancellation effects, should be easily seen at the
precision level attained for these space missions \citep{Balona1999}. 
Indeed, this expectation appears to have been fulfilled in {\it CoRoT}
observations of HD~50844 \citep{Poretti2009}.  This star appears to have a
very high mode density over the whole frequency range, indicating that modes
of relatively high $l$ are seen.  However, in a study of {\it Kepler} 
$\delta$~Sct stars, \citet{Balona2011c} find that, in general, the mode 
density is quite moderate and that HD~50844 is probably an exception.

In order to study mode density, one needs to be sure that frequencies are
correctly extracted from the data.  This is not an issue for the high
amplitude peaks in the periodogram, but since the number of modes is
expected to increase with decreasing amplitude, one needs to estimate the
probability that a given peak is due to noise (the false-alarm probability). 
The significance level of a frequency can be estimated easily only for the 
case of an equally-spaced time series, when the Lomb-Scargle criterion can 
be used \citep{Scargle1982}.  When the sampled times are not uniformly spaced, 
the problem is intractable and can only be solved by numerical means 
\citep{Frescura2008}.  Ground-based data are seldom equally spaced and the 
``four-sigma'' rule is often used to estimate the significance level 
\citep{Breger1993}.  This rule states that a peak is  deemed significant if 
its amplitude exceeds the background noise level of the periodogram by a 
factor of four.  This is a purely empirical rule with no statistical 
foundation, but does seem to be reasonable in many cases
\citep{Kuschnig1997, Koen2010}.  The {\it CoRoT} data are uniformly spaced 
and it is therefore important to apply the correct statistics.

 For a time series of duration, $t$, the width of a peak in the 
power spectrum is inversely proportional to $t$.  Thus for a time series of a 
given duration, there is a maximum limit on the number of peaks per frequency 
interval that can be measured. Since the mode density is expected to
increase with decreasing amplitude, blending of peaks due to the finite 
frequency resolution becomes very important. Thus the number of frequencies 
that can be extracted, and hence the mode density, depends not only on a
correct estimation of the false alarm probability, but also on effects
related to the finite frequency resolution.  These effects have not been 
taken into account in estimates of mode density by \citet{Poretti2009} 
and \citet{Balona2011c}.

From the above considerations, it is clear that one needs to carefully reconsider 
the question of significance in frequency extraction for {\it CoRoT} and {\it
Kepler} data and, in particular, the importance of frequency resolution and
close frequencies.  In this paper we use simulations at various mode
densities to determine the effect of mode density on frequency extraction.
We also compare results obtained with the Lomb-Scargle false alarm
probability (FAP) criterion with those calculated using the four-sigma 
rule.  It turns out that neither of these criteria are useful if the
number of modes increases with decreasing amplitude.  Finally, we compare
the amplitude density in the periodogram of HD\,50844 with those of {\it
Kepler} $\delta$~Scuti stars.

\section{The data and noise properties}

The {\it CoRoT} observations of HD~50844 consist of $n = 155827$ white light 
data points obtained between HJD~2452590.0684 and HJD~2452647.7817 (57.7~d) 
with a cadence of 30~s.  The time series is almost uninterrupted, except for 
data taken at the southern magnetic anomaly and some other points which were
rejected because they were clearly anomalous.

We first of all need to understand the noise properties of these data.  One 
way to do this is to select stars which appear to be constant, or vary the 
least.  The noise level in the periodogram of such a star may be used to 
calibrate the noise level in any other star in a purely empirical way.  

The definition of what is meant by the noise level in the periodogram needs
to be clarified.  In visual estimates of the noise level, this is generally
taken as the height of the majority of peaks.  The peaks in the periodogram
may be likened to blades of grass on a lawn and the mean height of the peaks
may be called the ``grass'' level.  This loose definition has been placed on 
a firmer footing by \citet{Balona2011a} who defined $\sigma_G = 2.5\sigma$, 
where $\sigma$ is the median height of the periodogram peaks in a region
free of signals.  In the four-sigma rule, a peak with amplitude $A$ may be 
considered significant if $A > 4\sigma_G$.  One may also define the mean 
background noise level as just the average of the power or the amplitude.
Since the background noise level typically increases towards low frequencies, 
and since the backgound is often difficult to estimate in crowded regions of 
the spectrum, the exact meaning of ``mean background level'' is often not 
clear. 

It can be shown that the mean amplitude, $A$, of the periodogram of pure
white noise with variance $\sigma_0$ is given by $A = 2\sigma_0/\sqrt{n}$, 
where $n$ is the number of points in the time series (see \citet{Kendall1976}).  
From the high-frequency tail of the periodogram of HD~50844 we find $A =
0.0035$~mmag which gives $\sigma_0 = 0.691$~mmag.  Cast in terms of apparent 
magnitude, $V$, and duration of the time series, $\Delta t$, the noise level 
in the periodogram can be written as
$$\log \sigma_G = a + \frac{1}{5}V - \frac{1}{2}\log \Delta t,$$  
where $a$ is a constant.  By fitting 175 of the least variable stars in the  
A--F range in the {\it Kepler} data, \citet{Balona2011a} found 
$$\log \sigma_G = -0.93 + 0.21V - 0.47\log \Delta t,$$ 
where $\sigma_G$ is measured in ppm and $\Delta t$ in days, confirming the 
expected relationship.  

\begin{figure}
\centering
\includegraphics{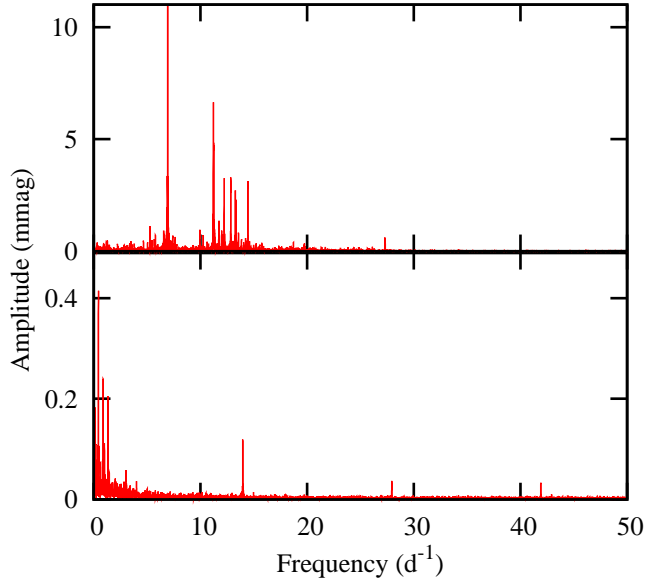}
\caption{Periodograms of HD~50844 (top) and HD~292790 (bottom).  The
estimated noise level for HD~50844 is $\sigma_G = 4.0$\,ppm while for
HD~292790 it is $\sigma_G = 5.0$\,ppm (see Fig.\,\ref{pnoise} for an expanded
view). }
\label{period}
\end{figure}

\begin{figure}
\centering
\includegraphics{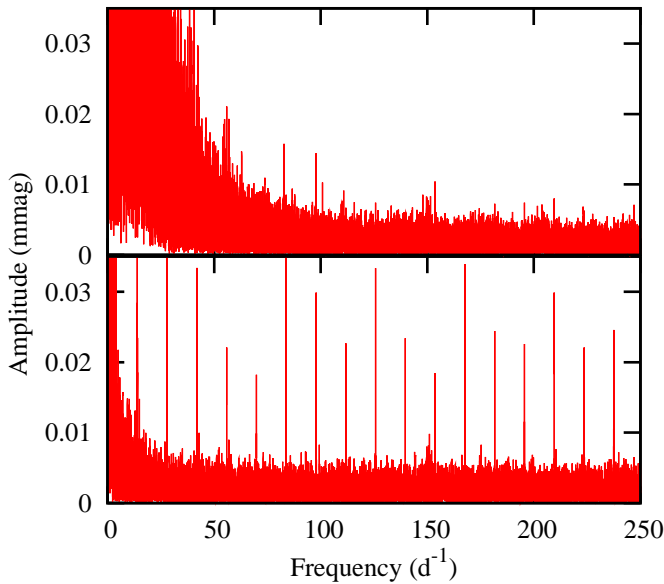}
\caption{Periodograms of HD~50844 (top) and HD~292790 (bottom) with expanded
amplitude scale to show the noise level.}
\label{pnoise}
\end{figure}

In the case of the {\it CoRoT} data, we do not have access to a large number
of constant or nearly constant A--F stars for a completely independent
estimate of the noise level.  However, we do have data for HD~292790, also 
measured by \citet{Poretti2009}.  This star is variable at low frequencies, 
but the noise level is well defined at the higher frequencies relevant for 
$\delta$~Sct stars.  The magnitude of HD~292790 is $V = 9.48$ and the 
time-series duration $\Delta t = 54.66$~d.  For the $\delta$~Scuti star 
HD~50844, $V = 9.09$ and $\Delta t = 57.70$~d.  Fig.~\ref{period}
shows the periodograms for these two stars.  Fig.~\ref{pnoise} shows an
expanded region to show the noise levels.  

The periodogram of HD~292790 can be understood in terms of rotation of a 
spotted star with frequency $f = 0.4386$~d$^{-1}$ \citep{Poretti2009}.  
The periodogram contains a signal at 13.9668~d$^{-1}$ and its harmonics 
which is the orbital frequency of the satellite.  For this star we 
estimate $\sigma_G = 5$ ppm, which allows us to determine the constant 
$a$ in the above equation.  We  obtain  
$$\log \sigma_G = -0.33 + 0.2V - 0.5\log \Delta t,$$ 
which means that {\it Kepler} data are about four times more precise than
{\it CoRoT} data.  This formula gives $\sigma_G = 4.0$~ppm for HD~50844,
which is indeed the noise level of the high frequency tail in Fig.~\ref{pnoise}.  
Thus, according to the four sigma rule, we may assume that for HD~50844 only 
frequencies  with amplitudes exceeding approximately 16~ppm are likely to be real.  

\begin{figure}
\centering
\includegraphics{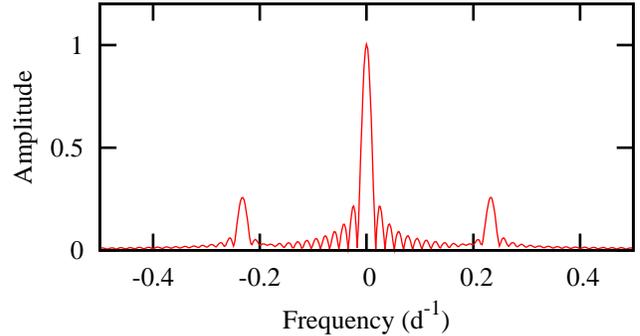}
\caption{Spectral window of the HD~50844 {\it CoRoT} data.}
\label{specwin}
\end{figure}

The spectral window of the {\it CoRoT} data for HD~50844 is shown in
Fig.~\ref{specwin}. The FWHM of the central peak is about 0.02~d$^{-1}$,
which means that frequencies separated by less than about 0.01~d$^{-1}$ are
not likely to be resolved.

\section{Tests of significance}

Although the four-sigma test of significance is widely used because it is 
simple to apply, it is by no means a rigorous test.  For example,
\citet{Koen2010} finds that the actual significance levels corresponding to 
the four sigma limit may vary by orders of magnitude, depending on the exact 
data configuration.  He finds that the number and time spacing of the
observations have little influence on the significance levels.  On the other
hand, the total duration of the time series, the frequency range that is
searched and previous prewhitening of the data greatly affect the significance
level of the four sigma rule. In particular, prewhitening removes too much
power at the given frequency, hence the estimated mean noise level is lower
than it should be, decreasing the false alarm probability.  This means that
repeated application of the rule on successively prewhitened data will tend
to assign significance to peaks which are probably noise.  \citet{Koen2010} 
also find that the four sigma rule, when applied to a single frequency peak,
is likely to underestimate its significance.  In other words, there are
peaks of lower amplitude which are significant, but which are considered to
be noise by the four-sigma rule.

More precise significance tests for periodograms have been discussed by, 
among others, \citet{Scargle1982, Horne1986, Koen1990} and 
\citet{Schwarzenberg1998}.  Somewhat different methods are discussed by 
\citet{Reegen2007, Reegen2008} and \citet{Baluev2008}.  An excellent discussion 
of the problem is presented by \citet{Frescura2008}.  The problem is clearly a 
very difficult one for data which is not uniformly sampled.  For uniformly-sampled 
data Scargle's significance test \citep{Scargle1982} is easy to calculate and
gives the false-alarm probability of a periodogram peak given the power
level and the noise variance of the data.  Since the {\it CoRoT} and {\it Kepler}
data are equally sampled, except for occasional small gaps, it is clear that
this test is to be preferred to the four-sigma rule.

For equally-spaced data, \citet{Scargle1982} shows that the false-alarm 
probability (FAP), $P(z)$, of a peak with power $z$ is given by
$$P(z) = 1 - \left\{1 - \exp(-z/\sigma^2_0)\right\}^{N_i},$$
where $\sigma^2_0$ is the noise variance of the data and $N_i$ is the number of
independent frequencies.  We call this the Lomb-Scargle significance
criterion.  It should be noted that in some references the power is normalized 
by the noise variance, while in others it is not.  We use the un-normalized definition.  
This relationship can be inverted to give the power for any FAP,
$$z = -\sigma^2_0\ln\left\{1 - \left(1 - P(z)\right)^{1/N_i}\right\}.$$
The number of independent frequencies is well defined only for
equally-spaced data and is given by $N_i = n/2$, where $n$ is the number of
data points \citep{Frescura2008}.  Astronomers usually prefer the periodogram 
to be a function of amplitude rather than power.  For our definition of power, 
$z$, the amplitude is given by $A = 2\sqrt{z/n}$.  Given a certain FAP,
$P_A$, one may then calculate an amplitude, $A_A$, which corresponds to this 
probability,
$$A_A = 2 \sqrt{ -\frac{\sigma^2_0}{n}\ln\left\{1 - \left(1 -
P_A\right)^{1/N_i}\right\}}.$$
For {\it CoRoT} observations of HD~50844, we find $A_A = 0.014$~mmag for
$P_A = 0.01$ and $A_A = 0.015$~mmag for $P_A = 0.001$.  The four-sigma rule
suggests that only amplitudes above 0.016~mmag are significant, which is
certainly true, but is very stringent and corresponds to a FAP of less than
0.001, whereas a FAP of $P_A = 0.01$ is often sufficient.

\section{Simulations}

Before we apply the Lomb-Scargle criterion to the {\it CoRoT} data of HD~50844, 
it is important to test it on synthetic data.  These tests take the actual 
times of {\it CoRoT} observations of this star to generate a time series 
comprising of many sinusoidal variations with randomly generated
frequencies, amplitudes and phases.  The simulations consist of uniformly
distributed frequencies in the range 0--30~d$^{-1}$, which is roughly the
frequency range of pulsations in HD~50844.  In $\delta$~Sct stars, the number of
modes increases sharply with decreasing amplitude.  To roughly mimic this,
the amplitudes in our simulations are exponentially distributed.  A noise error 
with a Gaussian distribution and standard deviation of 0.500~mmag was added to 
each point of the synthetic time series.  This leads to a mean periodogram
height of 0.0025~mmag or a grass noise level $\sigma_G = 0.006$~mmag,
roughly similar to that found in HD~50844.  The aim of these simulations is 
to determine the effect of mode density on frequency extraction and to test
the reliability of these frequencies in data with a high signal-to-noise (S/N) 
ratio.  We also wish to investigate the efficiency of the four-sigma rule and 
the Lomb-Scargle significance criterion.

The time series was analyzed using the Lomb periodogram and the algorithm 
described by \citet{Press1989} for fast implementation.  The peak of maximum
amplitude is located and its frequency measured by fitting a quadratic to
points around maximum amplitude.  This frequency, together with up to 20
previous frequencies, is used in a simultaneous least-squares Fourier fit to
the data.Once the limit of 20 frequencies has been reached, the original time 
series is replaced by the prewhitened time series and the process repeated
until the peak of highest amplitude is no longer significant.  
 
We start with the sparsest frequency set of 30 frequencies.  In this case
we are able to extract 65 frequencies for which FAP $>$ 0.01 (or 47
frequencies using the four-sigma limit) even though the simulated data has 
only 30 frequencies.  Of these frequencies, the first 28 of highest
amplitude match the known frequencies.  The two missing frequencies do
appear, but are far down on the list since they have very low amplitudes.  
The question that needs to be asked is where do the additional 35 frequencies
which have significant amplitudes come from?

\begin{figure}
\centering
\includegraphics{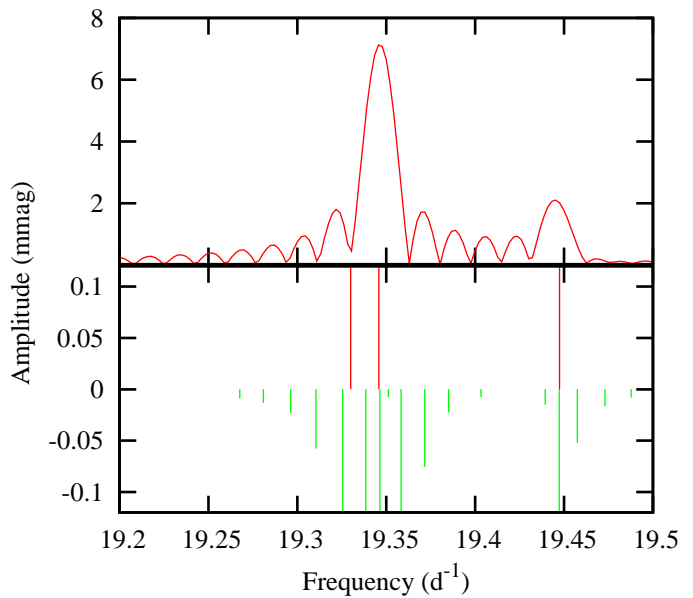}
\caption{Schematic periodogram of known frequency components (with positive
amplitudes) and extracted components (negative amplitudes) in a simulation.}
\label{a30}
\end{figure}

Fig.\,\ref{a30} shows the periodogram in a specific frequency region and also
a schematic periodogram of the same region which shows that the large peak
actually consists of two unresolved closely-spaced frequencies.  These  
frequencies and amplitudes are  $f_1 = 19.3458$~d$^{-1}$, $A_1 = 7.118$~mmag, 
$f_2 = 19.3300$~d$^{-1}$, $A_2 = 0.574$~mmag.  The figure also shows a
schematic periodogram of the extracted frequencies.  Instead of extracting 
just a single frequency, the code finds numerous, roughly equally-spaced
frequency components of relatively high amplitudes.  A third frequency at 
$f_3 = 19.4475$~d$^{-1}$, $A_3 = 1.855$~mmag also shows fictitious 
components.

The origin of these fictitious components are easy to understand.  They come
about because the extracted frequency of the unresolved peak is
19.3465~d$^{-1}$, which differs from the true frequency.  Even though the
difference is only 0.0007~d$^{-1}$, the prewhitened data still contains
significant signal because the amplitude of the unresolved peak is so high. 
Prewhitening, in fact, is equivalent to adding a fictional signal to the
data.  When the prewhitening frequency and amplitude is sufficiently close 
to the true values, both are removed through destructive interference.  If the
prewhitening frequencies and amplitudes not quite correct, what remains is a
sequence of equally-spaced frequencies of diminishing amplitude (an
interference pattern).  If the frequency of interest has a low amplitude,
the residual interference pattern may be below the noise level, which is
nearly always the case in ground-based observations.  Space data have such
high signal-to-noise ratio that the amplitudes of the residual interference
pattern is well above the noise level.  Thus far more frequencies are
extracted than actually exist.  Provided that the S/N level is sufficiently
high, successive frequency extraction will lead to an ever increasing number
of frequencies at low amplitudes which do not actually exist.

\begin{table}
\centering
\caption{Results of frequency extraction using the Lomb periodogram on
simulated data. Random frequencies uniformly distributed in the range 
0--30~d$^{-1}$ and random amplitudes exponentially distributed in the 
range 0--10~mmag were used. $N_0$ is the number of generated frequencies,
$N$ is the number of these frequencies with amplitudes above the FAP = 0.01 
threshold.  $N_{\rm ex}$ is the number of extracted frequencies with 
FAP $<$ 0.01 using the Lomb-Scargle periodogram.  The frequencies extracted 
using the four-sigma criterion is given by $N_4$.  The two last columns give 
the numbers of significant frequencies using non-linear optimization.  
$L_{\rm ex}$ refers to the numbers using the Lomb-Scargle FAP and 
$L_4$ to the four-sigma criterion.}
\begin{tabular}{rrrrrr}
\hline
\hline
 $N_0$ & $N$ & $N_{\rm ex}$& $N_4$ & $L_{\rm ex}$  & $L_4$ \\
\hline
  30 &   30 &   65 &   47 &   30 &   29 \\
  50 &   50 &  137 &   87 &   57 &   54 \\
 100 &   98 &  220 &  150 &  132 &  112 \\
 200 &  198 &  537 &  350 &  402 &  320 \\
 300 &  298 &  879 &  585 &  729 &  596 \\
 500 &  493 & 1611 & 1141 & 1439 & 1197 \\
 700 &  693 & 2101 & 1545 & 1976 & 1687 \\
1000 &  998 & 3711 & 2215 & 2566 & 2283 \\
2000 & 1982 & 3405 & 2805 & 3108 & 2797 \\
3000 & 2969 & 3599 & 2968 & 3488 & 3187 \\
\\
\hline
\end{tabular}
\label{syn}
\end{table}

Results of simulations using an increasing density of frequencies are shown 
in Table~\ref{syn}.  We note that the four-sigma criterion is more stringent
than the Lomb-Scargle FAP criterion, though it is clear that these criteria
are irrelevant in identifying the true frequencies.  Note also that the
number of extracted frequencies increases only slowly for $N > 1000$.  The
reason for this is due to the finite resolution imposed by the limited
duration of the time series.

There are two problems that come to light.  The first problem consists in
unavoidable errors in blended frequencies with high amplitudes, resulting in 
a cascade of low-amplitude (but highly significant) fictitious frequencies. 
The second problem is one of frequency resolution which increases the number
of blended frequencies and compounds the problem.  It might be possible to
recognize close frequencies using nonlinear optimization.  To test this
possibility we used the Levenberg-Marquard algorithm in combination with the
Lomb periodogram to optimize the frequencies, grouping closely-spaced
frequencies in a simultaneous optimization scheme.  Results are shown in
Table\,\ref{syn}.  While nonlinear optimization gives much better results
for low mode densities, it still fails badly as the density increases. 

Another symptom which arises from continuous prewhitening of data with very
high S/N is the development of a plateau in the periodogram consisting of a 
very large number of blended peaks of almost equal amplitude.  An example of
this can be seen in the simulations arising from the simulation with $N_0 =
1000$ frequencies.  Such a plateau occurs in other simulations, but is most
prominent in those with a high mode density.  The development of a plateau
can be seen in Fig.\,3 of \citet{Poretti2009}.

\begin{figure}
\centering
\includegraphics{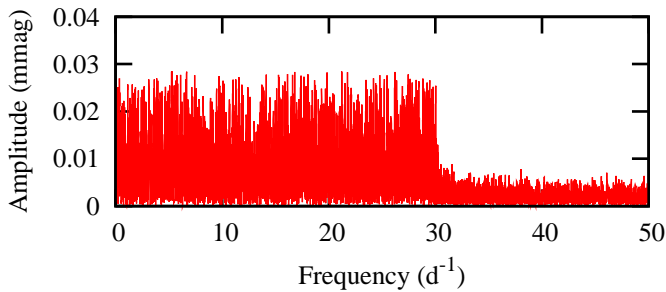}
\caption{Periodogram of simulated data with 1000 actual frequencies after
prewhitening by 2000 frequencies showing the characteristic plateau.}
\label{plateau}
\end{figure}

From these simulations we conclude that one should be very careful in
extracting frequencies with small amplitudes in stars with very high S/N data
containing several high-amplitude peaks.  In particular, no trust should be
placed in the very large number of frequencies  which results from
successive prewhitening in such stars.  This is undoubtedly the case in
HD\,50844.

\section{Amplitude distribution}

The simulations described above show that the effect of finite frequency
resolution is very important.  In fact, it is likely to be of far greater
importance well before the amplitude threshold appropriate to a given false
alarm probability.  A question that is directly relevant to estimation of
the mode density is the distribution of amplitudes.  In order to estimate
the mode density for the lowest detectable amplitudes, one needs to
calculate the amplitude distribution, i.e., the number of modes within a
given amplitude range.  From the results of the previous section, it is
clear that repeated prewhitening will lead to an ever increasing number of
false modes for low amplitudes.  In this section we investigate the
amplitude distribution using simulations.  

For this purpose, we generated a number of time series with random 
frequencies uniformly distributed in the frequency range 0--30~d$^{-1}$ 
with uniformly distributed amplitudes in the amplitude range 0--10~mmag.  
This differs from the analysis in the previous section where we used an 
approximately exponential amplitude distribution.  A uniform amplitude 
distribution is not expected in $\delta$~Sct stars, but it allows the 
known and extracted amplitude distributions to be more easily compared.  

Table\,\ref{uni} shows the number of simulated frequencies and the number 
of extracted frequencies using the Lomb periodogram.  We did not attempt to
apply non-linear optimization of the frequencies, since this does not
resolve the frequency resolution problem discussed in the previous section. 
Note that the number of extracted frequencies is much larger than the number
of real frequencies.  This is not surprising because, in general, the
amplitudes in this simulation are much larger than in the exponential
distribution which increases the cascading effect that arises when
frequencies are unresolved.

\begin{table}
\centering
\caption{Results of frequency extraction using the Lomb periodogram on
simulated data with uniform amplitude distribution.  $N_0$ is the total 
number of simulated frequencies and $N$ is the number of 
simulated frequencies with amplitudes above the FAP = 0.01 threshold.  
$N_{\rm ex}$ is the number of extracted frequencies with FAP $<$ 0.01 
using the Lomb-Scargle periodogram.}
\begin{tabular}{rrr}
\hline
\hline
   $N_0$ & $N$ & $N_{\rm ex}$ \\
\hline
  30 &     30 &   105 \\
  50 &     50 &   238 \\
 100 &     99 &   490 \\
 200 &    199 &  1226 \\
 300 &    299 &  1833 \\
 500 &    499 &  2684 \\
 700 &    699 &  3217 \\
1000 &    998 &  3711 \\
2000 &   1998 &  3711 \\
3000 &   2996 &  5212 \\
\\
\hline
\end{tabular}
\label{uni}
\end{table}

\begin{figure}
\centering
\includegraphics{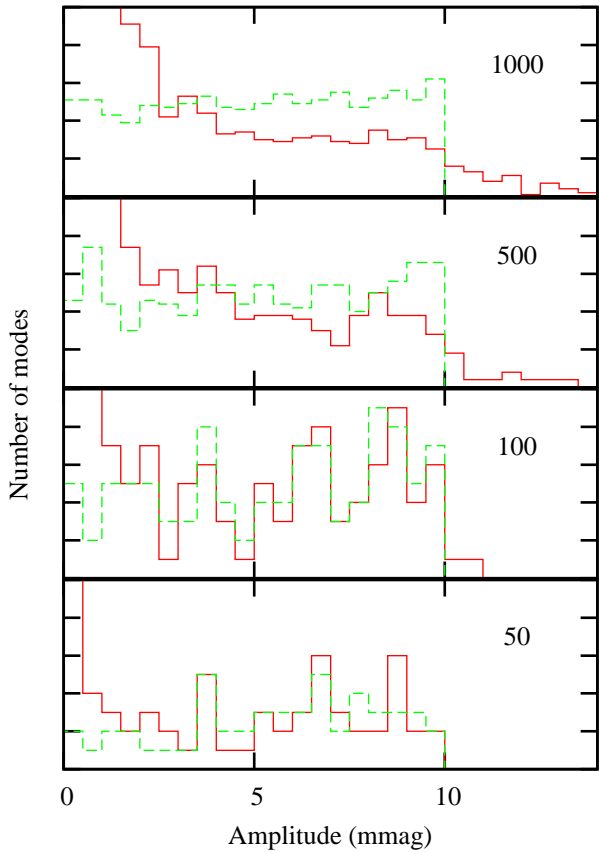}
\caption{Distribution of amplitudes extracted from the Lomb periodogram for
a selection of simulated data.  The solid histogram shows the relative number 
of extracted frequencies as a function of amplitude (mmags).  The dashed
histogram is the true distribution.  The panels are labeled according to the
total number of known frequencies.}
\label{amphist}
\end{figure}

\begin{table*}
\centering
\caption{Significant frequencies and amplitudes in HD~50844.  The frequency $f$ is 
in d$^{-1}$, the amplitude $A$ in mmillimags and the phase $\phi$ in radians.
This is a fit to $A\cos(2\pi f(t - t_0) + \phi)$ with $t_0 = 2590.0000$. 
The second and third columns list mode identifications from spectroscopy,
$(l,m)$, and radial velocity amplitudes, $A_{\rm RV}$ (km\,s$^{-1}$) from
\citet{Poretti2009}.}
\begin{tabular}{lrrrrr}
\hline
\hline
 Term   & (l,m)  & $A_{\rm RV}$   &\multicolumn{1}{c}{$f$ (d$^{-1}$)}
&\multicolumn{1}{c}{$A$ (mmag)}  & \multicolumn{1}{c}{$\phi$ (rad)}\\ 
\hline
$f_1 $   & (0,0) & 0.72 & $ 6.92527 \pm 0.00007$ & $16.73 \pm 0.01$ & $ 0.575 \pm 0.001$ \\
$f_2 $   & (3,1) & 0.39 & $11.21672 \pm 0.00008$ & $ 6.76 \pm 0.01$ & $ 1.559 \pm 0.001$ \\
$f_3 $   & (5,1) &      & $11.25803 \pm 0.00010$ & $ 4.37 \pm 0.01$ & $ 0.773 \pm 0.002$ \\
$f_4 $   & (3,3) & 0.52 & $12.84848 \pm 0.00010$ & $ 3.55 \pm 0.01$ & $-2.281 \pm 0.002$ \\
$f_5 $   &       & 0.45 & $12.23840 \pm 0.00009$ & $ 3.53 \pm 0.01$ & $-0.010 \pm 0.002$ \\
$f_6 $   & (4,3) &      & $14.44681 \pm 0.00008$ & $ 3.08 \pm 0.01$ & $ 1.986 \pm 0.003$ \\
$f_7 $   & (3,2) &      & $13.27339 \pm 0.00008$ & $ 2.87 \pm 0.01$ & $-2.871 \pm 0.003$ \\
$f_8 $   & (5,3) & 0.11 & $13.35871 \pm 0.00008$ & $ 2.51 \pm 0.01$ & $ 1.803 \pm 0.003$ \\
$f_9 $   & (11,7)& 0.05 & $11.75098 \pm 0.00010$ & $ 1.58 \pm 0.01$ & $ 1.585 \pm 0.005$ \\
$f_{10}$ & (2,-2)& 0.08 & $ 5.26680 \pm 0.00011$ & $ 1.24 \pm 0.01$ & $ 1.362 \pm 0.007$ \\
$f_{11}$ &       & 0.20 & $14.46124 \pm 0.00011$ & $ 1.24 \pm 0.01$ & $ 0.071 \pm 0.007$ \\
$f_{12}$ & (4,2) & 0.10 & $14.43221 \pm 0.00011$ & $ 1.08 \pm 0.01$ & $-2.167 \pm 0.008$ \\
$f_{13}$ &       & 0.10 & $ 9.95266 \pm 0.00011$ & $ 1.02 \pm 0.01$ & $ 2.798 \pm 0.008$ \\
         &       &      & $ 0.00497 \pm 0.00011$ & $ 1.26 \pm 0.04$ & $-2.483 \pm 0.029$ \\
$f_{14}$ &       &      & $11.98370 \pm 0.00011$ & $ 0.94 \pm 0.01$ & $ 1.148 \pm 0.009$ \\
$f_{15}$ &       &      & $ 6.55652 \pm 0.00011$ & $ 0.82 \pm 0.01$ & $ 2.895 \pm 0.010$ \\
$f_{17}$ &       &      & $13.56876 \pm 0.00011$ & $ 0.85 \pm 0.01$ & $-3.090 \pm 0.010$ \\
$f_{16}$ &       &      & $ 7.40484 \pm 0.00011$ & $ 0.85 \pm 0.01$ & $ 1.412 \pm 0.010$ \\
$f_{18}$ & (5,0) &      & $10.26038 \pm 0.00011$ & $ 0.79 \pm 0.01$ & $-0.864 \pm 0.010$ \\
$f_{19}$ &       & 0.08 & $ 5.78209 \pm 0.00011$ & $ 0.74 \pm 0.01$ & $-1.964 \pm 0.011$ \\
         &       &      & $11.70934 \pm 0.00012$ & $ 0.69 \pm 0.01$ & $-1.332 \pm 0.012$ \\
$f_{20}$ &       &      & $ 6.62851 \pm 0.00012$ & $ 0.71 \pm 0.01$ & $ 1.919 \pm 0.012$ \\
         &       &      & $12.12271 \pm 0.00012$ & $ 0.70 \pm 0.01$ & $ 1.160 \pm 0.012$ \\
         &       &      & $ 7.63511 \pm 0.00012$ & $ 0.66 \pm 0.01$ & $ 2.999 \pm 0.013$ \\
         &       &      & $27.29546 \pm 0.00012$ & $ 0.66 \pm 0.01$ & $ 2.046 \pm 0.013$ \\
$f_{22}$ & (3,1) &      & $14.47801 \pm 0.00011$ & $ 0.68 \pm 0.01$ & $ 1.201 \pm 0.013$ \\
         &       &      & $14.23452 \pm 0.00011$ & $ 0.56 \pm 0.01$ & $ 2.125 \pm 0.015$ \\
$f_{27}$ & (4,-2)&      & $ 5.67451 \pm 0.00012$ & $ 0.61 \pm 0.01$ & $-1.642 \pm 0.014$ \\
$2f_1$   &       &      & $13.84992 \pm 0.00012$ & $ 0.59 \pm 0.01$ & $-2.488 \pm 0.014$ \\
         &       &      & $ 4.66075 \pm 0.00012$ & $ 0.53 \pm 0.01$ & $ 0.390 \pm 0.016$ \\
         &       &      & $ 1.28065 \pm 0.00012$ & $ 0.51 \pm 0.01$ & $ 3.041 \pm 0.016$ \\
$f_{30}$ &    -  & 0.08 & $ 5.41839 \pm 0.00012$ & $ 0.53 \pm 0.01$ & $-2.480 \pm 0.016$ \\
$f_{32}$ & (5,-2)&      & $ 5.49082 \pm 0.00012$ & $ 0.51 \pm 0.01$ & $-0.946 \pm 0.016$ \\
         &       &      & $ 3.48107 \pm 0.00012$ & $ 0.47 \pm 0.01$ & $ 0.403 \pm 0.018$ \\
         &       &      & $11.33797 \pm 0.00012$ & $ 0.48 \pm 0.01$ & $-0.782 \pm 0.017$ \\
         &       &      & $ 1.17976 \pm 0.00012$ & $ 0.46 \pm 0.01$ & $-2.041 \pm 0.018$ \\
$f_{44}$ & (6,4) &      & $14.60065 \pm 0.00012$ & $ 0.46 \pm 0.01$ & $ 0.842 \pm 0.018$ \\
         &       &      & $14.26363 \pm 0.00012$ & $ 0.46 \pm 0.01$ & $ 0.472 \pm 0.018$ \\
         &       &      & $18.72302 \pm 0.00012$ & $ 0.45 \pm 0.01$ & $ 1.073 \pm 0.019$ \\
         &       &      & $ 7.25400 \pm 0.00012$ & $ 0.44 \pm 0.01$ & $ 2.297 \pm 0.019$ \\
         &       &      & $10.62982 \pm 0.00012$ & $ 0.43 \pm 0.01$ & $-1.538 \pm 0.019$ \\
         &       &      & $14.06179 \pm 0.00012$ & $ 0.42 \pm 0.01$ & $ 2.927 \pm 0.020$ \\
         &       &      & $15.75242 \pm 0.00012$ & $ 0.42 \pm 0.01$ & $-0.870 \pm 0.020$ \\
$f_{43}$ & (4,-2)& 0.12 & $ 5.04317 \pm 0.00012$ & $ 0.42 \pm 0.01$ & $-1.560 \pm 0.020$ \\
         &       &      & $ 0.32135 \pm 0.00012$ & $ 0.42 \pm 0.01$ & $-0.776 \pm 0.020$ \\
         &       &      & $13.06176 \pm 0.00013$ & $ 0.40 \pm 0.01$ & $ 3.040 \pm 0.021$ \\
$f_{50}$ &(12,10)&      & $15.22420 \pm 0.00012$ & $ 0.39 \pm 0.01$ & $-2.349 \pm 0.021$ \\
$f_{46}$ &(14,12)&      & $19.75031 \pm 0.00012$ & $ 0.39 \pm 0.01$ & $-0.179 \pm 0.021$ \\
         &       &      & $ 1.23137 \pm 0.00013$ & $ 0.38 \pm 0.01$ & $-1.064 \pm 0.022$ \\
         &       &      & $ 3.38306 \pm 0.00013$ & $ 0.37 \pm 0.01$ & $ 2.196 \pm 0.023$ \\
         &       &      & $ 0.93230 \pm 0.00013$ & $ 0.37 \pm 0.01$ & $ 0.115 \pm 0.023$ \\
$f_{51}$ & (8,5) & 0.08 & $11.64339 \pm 0.00013$ & $ 0.36 \pm 0.01$ & $ 1.352 \pm 0.023$ \\
         &       &      & $ 2.22293 \pm 0.00013$ & $ 0.35 \pm 0.01$ & $-2.960 \pm 0.024$ \\
         &       &      & $ 3.79566 \pm 0.00013$ & $ 0.34 \pm 0.01$ & $-0.829 \pm 0.024$ \\
         &       &      & $ 3.57994 \pm 0.00013$ & $ 0.33 \pm 0.01$ & $-0.908 \pm 0.025$ \\
         &       &      & $ 6.57856 \pm 0.00013$ & $ 0.35 \pm 0.01$ & $ 1.095 \pm 0.024$ \\
         &       &      & $15.56830 \pm 0.00013$ & $ 0.33 \pm 0.01$ & $ 2.248 \pm 0.025$ \\
         &       &      & $ 2.85138 \pm 0.00013$ & $ 0.32 \pm 0.01$ & $-1.230 \pm 0.026$ \\
         &       &      & $ 8.56364 \pm 0.00013$ & $ 0.31 \pm 0.01$ & $ 1.316 \pm 0.027$ \\
\\
\hline
\end{tabular}
\label{hd}
\end{table*}

In Fig.\ref{amphist} we show the distribution of amplitudes for four
selected simulations.  In all cases, the true distribution is flat (uniform
amplitude distribution) and extends from 0--30~d$^{-1}$.  The amplitude
distributions derived from the extracted frequencies are very different and
always show a large number of low-amplitude frequencies (out of scale in the
Figure).  In addition, there is considerable spillage to high amplitudes
due to two or more frequencies which are unresolved.  The larger the number
of frequencies the greater the number of frequencies at very low and very
high amplitudes.

It is clear that high mode density leads to a grossly distorted amplitude
distribution obtained by successive prewhitening.  In fact, numerical
experiments show that if successive prewhitening is performed until the
false alarm probability of the highest peak reaches a reasonable level ($P_A
= 0.01 or 0.001$), the resulting amplitude distribution is practically the same,
independently of the actual distribution that is chosen for the simulation. 
The distribution is dominated by the low-amplitude peaks which are mostly
false peaks arising from unresolved frequencies.  

\section{Avoiding prewhitening}

Given the problems associated with continuous prewhitening, it is clear that
another method of extracting frequencies is desirable for data with very
high S/N.  One possibility is simply to locate the peaks in the periodogram
of the raw (i.e. non-prewhitened) data.  This has to be done with care
and in combination with prewhitening of the most dominant peaks because much
of the structure in the periodogram is a result of the window function.

\section{Frequencies in HD~50844}

If we apply the Lomb-Scargle criterion to the 30-s cadence data of HD~50844
we find that there are about 1800 frequencies with amplitudes in excess of 
0.014~mmag  ($P_A = 0.01$) and 1700 frequencies in excess of 0.015~mmag 
($P_A = 0.001$), mostly  within the range 0--30~d$^{-1}$.  This is
equivalent to 60 frequencies per d$^{-1}$ which is at the limit of
frequency resolution.  As we have seen, these numbers are likely to be
considerably larger than actually present in the star.

In Table\,\ref{hd} we list the frequencies of highest amplitude.  Since
all these frequencies have quite large amplitudes, there is no danger of
them being fictitious.  Our frequencies and amplitudes agree very well with 
those of \citet{Poretti2009} and we have adopted the same numbering scheme for
the frequencies.  There are very few combination frequencies; we find 
$f_{10} = 4f_1 - 2f_2$  and $f_{12} = 2f_4-f_3$, but these are most probably 
coincidences.  The harmonic $2f_1$ is clearly visible.

\section{Comparison with {\it Kepler} $\delta$~Scuti stars}

One question that is of interest is whether HD~50844 has a higher mode
density than other $\delta$~Scuti stars.  We cannot really answer this
question because it is not possible to extract frequencies below a certain
threshold level.  Also, we have seen that the finite time resolution and the
prewhitening technique provide severe limitations on the frequencies that
can be reliably extracted.  What can be answered is whether the amplitude
density in the periodogram is larger than in other $\delta$~Scuti stars.
To do this we need to compare the amplitude density in HD~50844 with those
of other stars.  

The {\it Kepler} data provide by far the most homogeneous sample of light 
curves of $\delta$~Sct stars.  Nearly all $\delta$~Sct stars observed by 
{\it Kepler} were discovered in the data from Quarters 0,1 and 2 which are 
in the public domain.  However, for most of these stars the length of the 
time series is about 30~d.   Since frequency resolution is important for
determining mode density, we need to ensure that the length of the time
series for each star is the same.  We decided to truncate the time series of
all the {\it Kepler} stars to 25~d in order to obtain the largest sample of
$\delta$~Scuti stars.  Also, we limit the data to stars observed with short
cadence (exposure times of about 60~s).  While considerably more {\it
Kepler} $\delta$~Sct stars have been observed with long cadence (exposure
time about 30~min), these cannot be used because the highest frequency
that can be detected is only about 24~d$^{-1}$, whereas most $\delta$~Sct
stars still have modes with frequencies as high as 50~d$^{-1}$.

\begin{figure}
\centering
\includegraphics{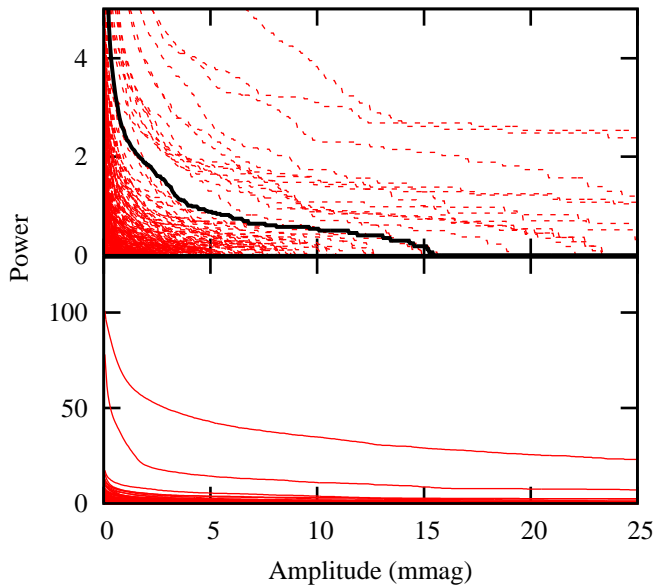}
\caption{Relative amplitude density as a function of amplitude for {\it Kepler}
$\delta$~Scuti stars.  The top panel shows the same plot expanded to show
HD~50844 (thick line).}
\label{moden}
\end{figure}

There are 357 $\delta$~Scuti stars observed in short-cadence mode in the
{\it Kepler} database.  Since all these stars were observed for the same
length of time, we can use the area of the periodogram above a certain 
amplitude level as proportional to the number of modes with amplitudes higher
than this level.  Fig.\,\ref{moden} shows this plot for the {\it Kepler}
$\delta$~Sct stars.  The top panel shows the plot for HD\,50844; the star
is clearly not exceptional in this regard.

\section{Conclusion}

Using simulations, we show that it is not possible to extract frequencies
reliably down to the expected noise level using successive prewhitening.  
The reason is due to unresolved peaks in the periodogram which become more 
numerous as the amplitude decreases.  Such unresolved peaks lead to a
multitude of erroneous frequencies. Prewhitening by a frequency which differs 
from the true frequency by a significant amount leaves many frequency residuals 
of lower amplitude.  In the high S/N data from space missions, these spurious 
residual frequencies often have significant amplitudes, leading to a cascading 
effect of spurious frequencies.  Simulations show that frequency extraction 
using the Lomb-Scargle or four sigma criterion can lead to a gross over-estimate of
the true number of frequencies.  Using nonlinear optimization does not solve
this problem.  It appears that the extraordinary high mode density of the
{\t CoRoT} star HD\,50844 found by \citet{Poretti2009} is not real, but a
result of this phenomenon.

Because of the effect described above, it is not possible to distinguish
between frequencies present in the star and spurious low-amplitude frequencies 
caused by prewhitening of unresolved frequency groups.  Thus it is 
not possible to count the number of individual modes with any degree of
certainty below a certain amplitude level.  One can, however, compare the
power in the periodogram above a given amplitude for different stars.  We
made such a comparison for HD\,50844 with 357 $\delta$~Scuti stars in the
{\it Kepler} public archive.  It turns out that HD\,50844 is not exceptional
in this regard.

\bibliographystyle{mn2e}
\bibliography{pgram}

\begin{thebibliography}{17}
\expandafter\ifx\csname natexlab\endcsname\relax\def\natexlab#1{#1}\fi

\bibitem[{{Balona}(2011)}]{Balona2011a}
{Balona} L.~A., 2011, \mnras, 415, 1691

\bibitem[{{Balona} \& {Dziembowski}(1999)}]{Balona1999}
{Balona} L.~A., {Dziembowski} W.~A., 1999, \mnras, 309, 221

\bibitem[{{Balona} \& {Dziembowski}(2011)}]{Balona2011c}
---, 2011, \mnras, 417, 591

\bibitem[{{Baluev}(2008)}]{Baluev2008}
{Baluev} R.~V., 2008, \mnras, 385, 1279

\bibitem[{{Breger} {et~al.}(1993){Breger}, {Stich}, {Garrido}, {Martin},
  {Jiang}, {Li}, {Hube}, {Ostermann}, {Paparo}, \& {Scheck}}]{Breger1993}
{Breger} M., {Stich} J., {Garrido} R., {Martin} B., {Jiang} S.~Y., {Li} Z.~P.,
  {Hube} D.~P., {Ostermann} W., {Paparo} M., {Scheck} M., 1993, \aap, 271, 482

\bibitem[{{Frescura} {et~al.}(2008){Frescura}, {Engelbrecht}, \&
  {Frank}}]{Frescura2008}
{Frescura} F.~A.~M., {Engelbrecht} C.~A., {Frank} B.~S., 2008, \mnras, 388,
  1693

\bibitem[{{Horne} \& {Baliunas}(1986)}]{Horne1986}
{Horne} J.~H., {Baliunas} S.~L., 1986, \apj, 302, 757

\bibitem[{{Kendall}(1976)}]{Kendall1976}
{Kendall} M., 1976, {Time-Series}. {Charles Griffin \& Co. Ltd.}, p. 104

\bibitem[{{Koen}(1990)}]{Koen1990}
{Koen} C., 1990, \apj, 348, 700

\bibitem[{{Koen}(2010)}]{Koen2010}
---, 2010, \apss, 329, 267

\bibitem[{{Kuschnig} {et~al.}(1997){Kuschnig}, {Weiss}, {Gruber}, {Bely}, \&
  {Jenkner}}]{Kuschnig1997}
{Kuschnig} R., {Weiss} W.~W., {Gruber} R., {Bely} P.~Y., {Jenkner} H., 1997,
  \aap, 328, 544

\bibitem[{{Poretti} {et~al.}(2009){Poretti}, {Michel}, {Garrido},
  {Lef{\`e}vre}, {Mantegazza}, {Rainer}, {Rodr{\'{\i}}guez}, {Uytterhoeven},
  {Amado}, {Mart{\'{\i}}n-Ruiz}, {Moya}, {Niemczura}, {Su{\'a}rez}, {Zima},
  {Baglin}, {Auvergne}, {Baudin}, {Catala}, {Samadi}, {Alvarez}, {Mathias},
  {Papar{\`o}}, {P{\'a}pics}, \& {Plachy}}]{Poretti2009}
{Poretti} E., {Michel} E., {Garrido} R., {Lef{\`e}vre} L., {Mantegazza} L.,
  {Rainer} M., {Rodr{\'{\i}}guez} E., {Uytterhoeven} K., {Amado} P.~J.,
  {Mart{\'{\i}}n-Ruiz} S., {Moya} A., {Niemczura} E., {Su{\'a}rez} J.~C.,
  {Zima} W., {Baglin} A., {Auvergne} M., {Baudin} F., {Catala} C., {Samadi} R.,
  {Alvarez} M., {Mathias} P., {Papar{\`o}} M., {P{\'a}pics} P., {Plachy} E.,
  2009, \aap, 506, 85

\bibitem[{{Press} \& {Rybicki}(1989)}]{Press1989}
{Press} W.~H., {Rybicki} G.~B., 1989, \apj, 338, 277

\bibitem[{{Reegen}(2007)}]{Reegen2007}
{Reegen} P., 2007, \aap, 467, 1353

\bibitem[{{Reegen} {et~al.}(2008){Reegen}, {Gruberbauer}, {Schneider}, \&
  {Weiss}}]{Reegen2008}
{Reegen} P., {Gruberbauer} M., {Schneider} L., {Weiss} W.~W., 2008, \aap, 484,
  601

\bibitem[{{Scargle}(1982)}]{Scargle1982}
{Scargle} J.~D., 1982, \apj, 263, 835

\bibitem[{{Schwarzenberg-Czerny}(1998)}]{Schwarzenberg1998}
{Schwarzenberg-Czerny} A., 1998, \mnras, 301, 831

\end{thebibliography}

\label{lastpage}

\end{document}